\documentclass[preprint, aps, amsmath, amssymb, nofootinbib]{revtex4-2}
\usepackage{graphicx}
\usepackage{subcaption}
\usepackage[colorlinks=true,linkcolor=blue,urlcolor=blue,citecolor=blue]{hyperref}
\def\Xint#1{\mathchoice
   {\XXint\displaystyle\textstyle{#1}}%
   {\XXint\textstyle\scriptstyle{#1}}%
   {\XXint\scriptstyle\scriptscriptstyle{#1}}%
   {\XXint\scriptscriptstyle\scriptscriptstyle{#1}}%
   \!\int}
\def\XXint#1#2#3{{\setbox0=\hbox{$#1{#2#3}{\int}$}
     \vcenter{\hbox{$#2#3$}}\kern-.5\wd0}}

\def\dashint{\Xint-}

\begin{document}

\title{Beth-Uhlenbeck equation for the thermodynamics of fluctuations in a generalized (2+1)D Gross-Neveu model}
\author{Biplab Mahato}
\email{biplab.mahato@uwr.edu.pl}
\affiliation{Institute of Theoretical Physics, University of Wroc\l{}aw, Plac Maxa Borna, 50-204 Wroc\l{}aw, Poland}
\author{David Blaschke}
\affiliation{Institute of Theoretical Physics, University of Wroc\l{}aw, Plac Maxa Borna, 50-204 Wroc\l{}aw, Poland}
\affiliation{Center for Advanced Systems Understanding (CASUS), Untermarkt 20, D-02826 G{\"o}rlitz, Germany}
\affiliation{Helmholtz-Zentrum Dresden-Rossendorf (HZDR), Bautzener Landstrasse 400, D-01328 Dresden, Germany}
\author{Dietmar Ebert}%
\affiliation{Institut f{\"u}r Physik, Humboldt Universit{\"a}t zu Berlin, Newtonstraße 15, 12489 Berlin, Germany}

\date{\today}

\begin{abstract}
We study a generalized version of the Gross-Neveu model in 2+1 dimensions. The model is inspired from Graphene, which shows a linear dispersion relation near the Dirac points. The phase structure and the thermodynamic properties in the mean field approximation have been studied before. Here, we go beyond the mean field level by deriving a Beth-Uhlenbeck equation for Gaussian fluctuations formulated in phase shift solutions, which we explore numerically, for the first time including their momentum dependence. We discuss the excitonic mass, fluctuation pressure, and phase shifts. The inclusion of momentum dependence in the phase shift shows a significant difference from the Lorentz-boosted version of the phase shift previously used in the literature. We find resurrection of the pseudoscalar bound states at large momentum above Mott temperature and show that the presence of Landau modes significantly contributes to the fluctuation pressure.
\end{abstract}
\maketitle

\section{Introduction}

Two-dimensional materials have been the center of attention in condensed matter physics for the last two decades. The field gained momentum after the discovery of Dirac points in graphene \cite{novoselovTwodimensionalGasMassless2005}. These are special points in the reciprocal lattice space where the conduction and valence bands touch each other, giving a linear dispersion relation to the electrons. The physics near these points is of particular importance as 
{they realize the case of chiral fermion systems at low energy}.
%this allows the access of Dirac physics 
For a brief review of the properties, see \cite{CastroNeto:2007fxn}; other interesting properties can be found in \cite{Gonzalez:2010jv,Hofmann:2014nwa} etc.

In this article, we consider a particular approximate theory that can be used to mimic the behavior near the Dirac points, namely, the Gross-Neveu model. The Gross-Neveu model was introduced in \cite{Gross:1974jv} for one spatial dimension to study the chiral symmetry breaking in quantum chromodynamics (QCD).  
For a discussion of the main aspects of this theory, including the phase structure, see \cite{Barducci:1994cb,Thies:2003kk,Ciccone:2022zkg}. 
Previously, a $(3+1)$ dimensional four-fermion model known as the Nambu-Jona-Lasinio (NJL) model \cite{Nambu:1961tp} was widely studied as a chiral effective theory. The model was initially formulated in terms of nucleons and mesons. It was then reformulated in terms of quarks in \cite{Eguchi:1976iz, Kikkawa:1976fe} and extensively studied in \cite{Volkov:1982zx,Volkov:1984kq, Hatsuda:1985eb, Hatsuda:1984jm}. 
{While the original version of the} model ignores the gluonic contribution,  
%which is slightly fixed by the inclusion of Polyakov loop 
{recent developments include a gluon background field in the Polyakov gauge} \cite{Fukushima:2003fw}. 
The 
%inclusion of the Polyakov loop has successfully 
{Polyakov-loop} improved NJL model has
been applied successfully to describe lattice QCD data at finite temperature \cite{Ratti:2005jh,Hansen:2019lnf}.

In this article, we consider the $(2+1)$ dimensional version of the Gross-Neveu model. 
The model is interesting as it has a nontrivial phase structure and can be solved exactly in the large $N$ limit. The model has been studied in the context of graphene in \cite{Ebert:2015hva, Zhukovsky:2015ncz}. These works explored the phase structure of this model in the mean field approximation. Recently, in Ref. \cite{Ebert:2018dzs} the authors used the Beth-Uhlenbeck approach introduced in \cite{Blaschke:2013zaa} for the $(3+1)$ dimensional case to go beyond the mean field. The paper focuses solely on the theoretical aspects and does not include any numerical calculations.  In this article, we bridge that gap and explore the model numerically in both the mean field and beyond the mean field approximations. We also include the finite momentum effect on the polarisation function and include an analytical formula for the imaginary part of the polarisation function.

The paper is organized as follows. In the next section, we introduce the model. 
Section \ref{sec:mfa} is dedicated to the application of the mean field approximation method to this model. It explores the phase diagram and thermodynamical quantities at the level of this approximation. In section \ref{sec:bu}, we introduce Beth-Uhlenbeck approach to go beyond the mean field. In this section, we describe excitonic bound states, phase shifts, pressure, and Landau damping. Finally, in the last section, we present our Conclusions.

\section{Model}
\label{sec:model}

The Lagrangian of the Gross-Neveu model consists of two parts $\mathcal{L} = \mathcal{L}_0+\mathcal{L}_{\text{int}}$. The free part is the Dirac term \footnote{Representation of the gamma matrices can be found in the appendix \ref{appendix:gamma}}
\begin{equation}
	\mathcal{L}_0 = \bar{\psi}(\gamma_\mu\partial_\mu  - \mu\gamma_3+m_0)\psi
\end{equation}
and the interaction part is the four-fermion local interaction of the form $-(\bar{\psi}\psi)^2$. In \cite{Ebert:2018dzs} the authors have introduced a generalized version of the model with four different couplings,
% \begin{equation}
% 	\mathcal{L}_{\text{int}} = -\left(\frac{G_1}{2N}(\bar{\psi}\psi)^2+\frac{G_2}{2N}(\bar{\psi}\gamma_{45}\psi)^2+\frac{H_1}{2N}(\bar{\psi}i\gamma_5\psi)^2+\frac{H_2}{2N}(\bar{\psi}i\gamma_4\psi)^2\right)
% \end{equation}

\begin{equation}
	\mathcal{L}_{\rm int} = -\sum_{i=1}^4 \frac{G_i}{2N}(\bar{\psi}\Gamma_i\psi)^2.
\end{equation}
where $\Gamma_i = \{I, \gamma_{45}, i\gamma_5, i\gamma_4\}.$ Appendix \ref{appendix:gamma} contains the representation of gamma functions used in the text.

Motivations to use such a model, as well as its basic properties, are considered in \cite{Ebert:2018dzs,Gusynin:2007ix}. In this article, we investigate the model numerically.

The power counting argument reveals that the model is not renormalizable in weak-coupling perturbation theory. It is shown that in the $1/N$ expansion scheme, this model can be renormalized in each order \cite{Rosenstein:1988pt}. Also, one can find a renormalizable theory by bosonizing this theory and adding a quartic interaction term in the scalar auxiliary fields. This model is known as the Gross-Neveu-Yukawa model \cite{Erramilli:2022kgp} and had previously been used to model two-dimensional Dirac systems, including graphene \cite{Mihaila:2017ble}.

\section{Mean Field Approximation}
\label{sec:mfa}

% \textcolor{red}{What is the justification for using mean field approximation?
% 	- One justification can be from graphene, which near Dirac points behaves non-perturbatively (coupling constant $\sim\!2.2$ for pristine graphene).
% }

The steps to obtain the mean field results for the model are explained in the paper \cite{Ebert:2018dzs}. 
Here, we briefly summarize them and present numerical results {as a prerequisite for considering the fluctuations beyond the mean field} 
in the following sections.

First, we perform the Hubbard-Stratonovich transformation by introducing auxiliary fields $\Phi_i = -G_i(\bar{\psi}\Gamma_i\psi)/N$.
%$\sigma_i$ and $\varphi_i$, defined by
% \begin{equation}
% 	\begin{pmatrix}
% 		\sigma_i \\ \varphi_i
% 	\end{pmatrix} = -\begin{pmatrix}
% 		G_i \\ H_i
% 	\end{pmatrix} \bar{\psi}\Gamma_i\psi,\quad \Gamma = \{\{I, \gamma_{45}\},\{ i\gamma_5, i\gamma_4\}\}.
% \end{equation}
%Then the mean field approximation is to approximate the auxiliary fields by their expectation values.
%\begin{equation}
%	\Phi_i \to \bar{\Phi}_i + \delta\Phi_i.
%\end{equation}
This article considers the case where all the couplings are equal, i.e. $G_i = G$, which respects the SU(2) symmetry of the Coulomb interaction. 
The article \cite{Zhukovsky:2015ncz} considers other possibilities at the mean-field level.
In such a case of equal couplings, the partition function, after integrating out the Fermionic degrees of freedom, has the following form \cite{Ebert:2018dzs}
\begin{equation}\label{eqn:partition_function}
	\mathcal{Z} = \int\prod_{i=1}^4\mathcal{D}\Phi_i \rm{exp}\left\{ -N\int d^3x \sum_{i=1}^4\left (\frac{\Phi_i^2}{2G} - \delta_{i1}\kappa\Phi_i\right ) + N \rm Tr\ln S^{-1}\right\},
\end{equation}
%\begin{equation}\label{eqn:potential}
%	\Omega(T, \mu) = N\sum_{i=1}^4\left(\frac{\bar{\Phi}_i^2}{2G} -\delta_{i1}\kappa \bar{\Phi}_1\right) - \frac{N}{\beta}\rm{Tr}\ln (\beta S^{-1}_{\rm{mf}}).
%\end{equation}
where $S^{-1} = \gamma_\mu\partial_\mu - \mu\gamma_3 + \sum_{i=1}^4\Gamma_i\Phi_i$. 
We have absorbed the mass term into the first auxiliary field $\Phi_1 \to \Phi_1 - m_0$ and defined $\kappa = m_0/G$. 
{
The mean field approximation consists in dropping the path integration over the auxiliary fields ${\Phi}_i$ and replacing them by those values $\bar{\Phi}_i$ which maximize the grand thermodynamic potential 
\begin{equation}
\label{eq:omega}
    \Omega(T,\mu) = -\frac{1}{\beta l^2}\ln\mathcal{Z},
\end{equation}
where $l^2$ is the area term and $\beta = 1/T$ is the inverse temperature.
Performing the mean field approximation simplifies Eq. \eqref{eq:omega} to }
\begin{multline} \label{eqn:omega}
	\Omega_{\rm mf}(T,\mu) = N\sum_{i=1}^4\left(\frac{\bar{\Phi}_i^2}{2G} -\delta_{i1}\kappa \bar{\Phi}_1\right) - N\sum_{k=1}^2\int\frac{d^2p}{(2\pi)^2}\left[E_p^{(k)} + \frac{1}{\beta}\ln\left(1+e^{-\beta(E_p^{(k)}+\mu)}\right) 
    \right. \\ \left. 
    + \frac{1}{\beta}\ln\left(1+e^{-\beta(E_p^{(k)}-\mu)}\right)\right],
\end{multline}
% \begin{multline}
% 	\Omega(T,\mu) = N\sum_{i=1}^2\left(\frac{\bar{\sigma}_k^2}{2G_i} + \frac{\bar{\varphi}_k^2}{2H_i}-\delta_{i1}\kappa \bar{\sigma}_1\right) - N\sum_{i=1}^2\int\frac{d^2p}{(2\pi)^2}\left[E_p^{(i)} + \frac{1}{\beta}\ln\left(1+e^{-\beta(E_p^{(i)}+\mu)}\right) \right. \\
% 	\left. + \frac{1}{\beta}\ln\left(1+e^{-\beta(E_p^{(i)}-\mu)}\right)\right]
% \end{multline}
where $E_p^{(k)} = \sqrt{p^2+M_k^2}$ and $M_{1,2} = \bar{\Phi}_2 \pm \sqrt{\bar{\Phi}_1^2+\bar{\Phi}_3^2+\bar{\Phi}_4^2}$.

The integral {over the vacuum term}
in the above equation is divergent. 
To renormalize the theory, we introduce a momentum cutoff $\Lambda$ as a regulator. 
The divergent term can then be absorbed into the bare coupling constant $G$. 
We introduce a renormalized coupling constant $g$ defined by the relation 
\begin{equation}
    \label{eqn:g}
    \frac{1}{g}  = \frac{1}{G} - \frac{\Lambda}{\pi}.
\end{equation}
{Then, the renormalized thermodynamic potential takes the form}
%$\dfrac{1}{G_i} \to \dfrac{1}{G_i} - \dfrac{\Lambda}{\pi} = \dfrac{1}{g_i}$ and $\dfrac{1}{H_i} \to \dfrac{1}{H_i} - \dfrac{\Lambda}{\pi} = \dfrac{1}{h_i}$. 
\begin{equation}\label{eqn:renorm_grand_potential}
	\Omega_{\rm mf}^{\rm ren}(T,\mu) = \Omega_{\rm mf, vac}^{\rm ren} - \frac{N}{\beta}\sum_{k=1}^{2}\int%^{|p|<\Lambda} 
    \dfrac{d^2p}{(2\pi)^2}\left[ \ln\left(1+ e^{-\beta(E_p^{(k)} + \mu)}\right) + \ln\left(1+ e^{-\beta(E_p^{(k)} - \mu)}\right)\right],
\end{equation}
where the vacuum term is
\begin{equation}
	\Omega_{\rm mf, vac}^{\rm ren} = N\sum_{i=1}^{4}\left(\dfrac{1}{2g}\bar{\Phi}_i^2 - \delta_{i1}\kappa\bar{\Phi}_1\right)  + \sum_{k=1}^2\dfrac{|M_k|^3}{6\pi}.
\end{equation}
% \begin{equation}
% 	\Omega_{vac}^{ren} = N\sum_{k=1}^{2}\left[\dfrac{1}{2g_k}\bar{\sigma}_k^2 + \dfrac{1}{2h_k}\bar{\varphi}_k^2 - \delta_{k1}\kappa\bar{\sigma}_1 + \dfrac{|M_k|^3}{6\pi}\right]
% \end{equation}
There are several unknown parameters in Eq. \eqref{eqn:renorm_grand_potential}. 
The renormalized coupling $g$, the momentum cutoff $\Lambda$, and the expectation values of the auxiliary fields. 
The latter can be obtained by requiring that they extremize the grand potential. 
The extremization conditions are also known as the gap equations,

\begin{equation}
	\frac{\partial \Omega^{\rm ren}_{\rm mf}(T, \mu)}{\partial \bar{\Phi}_i} = 0.
\end{equation}

% \begin{equation}
% 	\frac{\partial \Omega^{ren}}{\partial \bar{\sigma}_1} = 0 \qquad \frac{\partial \Omega^{ren}}{\partial \bar{\varphi}_1} = 0
% \end{equation}
These equations admit solutions only if $\bar{\Phi}_{2,3,4} = 0$. 
In that case, $M_k = \bar{\Phi}_1$ and from now on, we denote this value as $m$,
\begin{equation}\label{eqn:gap_equation}
    \frac{\partial \Omega_{\rm{mf, vac}}^{\rm{ren}}}{\partial m} + \int
    %^{|p|<\Lambda} 
    \frac{d^2 p}{(2\pi)^2} \frac{2m}{E_p}(f(E_p + \mu) + f(E_p - \mu)) = 0.
\end{equation}
where $f(x)=(1+e^{\beta x})^{-1}$ is the Fermi function. 
{The integrand represents a logarithmic derivative and therefore the integral can be evaluated with the result}
\begin{equation}\label{eqn:gap_equation_2}
	\frac{m}{\pi}\left(\frac{\pi}{g}+m+\frac{1}{\beta}\ln\left[1+2e^{-\beta m}\cosh(\beta \mu)+e^{-2\beta m}\right]\right) = \kappa.
\end{equation}
Our model does not fix the coupling constant. 
Instead, we can define a mass scale $M = {\pi}/{|g|}$ and present all our results in units of that scale.
%\textcolor{red}{Is there a way to connect some physical scale with $M$?}.
The gap equation \eqref{eqn:gap_equation_2} admits a positive solution for $m$ (the fermion condensate is negative \cite{Parazian:2023wmj}), if 
{$G>{\pi}/{\Lambda}$ and according to \eqref{eqn:g} the renormalized coupling $g$ is negative}. 
It is important to note that the numerical results are not sensitive to the momentum cutoff if it is much larger than $M$. 
This is in strong contrast with the (3+1)D version of the same model, where the cutoff plays a very important role, and one needs to fix this cutoff by matching some of the observables with experimental data. %\textcolor{red}{add References.}

% In the figure \ref{fig:kdependence} we have shown the dynamical mass as a function of the temperature, chemical potential and the chiral parameter $\kappa$.

% \begin{figure}
% 	\centering
% 	\begin{subfigure}{0.48\textwidth}
% 		\includegraphics[width=\textwidth]{plots/mudependencesigma1.pdf}
% 	\end{subfigure}
% 	\begin{subfigure}{0.48\textwidth}
% 		\includegraphics[width=\textwidth]{plots/kappadependencesigma1.pdf}
% 	\end{subfigure}
% 	\caption{Phase diagram of the Gross Neveu model in mean field approximation. On the left (right) we show the dependence of the critical line on chemical potential (chiral mass term). In both the graphs, the region below(above) the critical line corresponds to the phase with nonzero(zero) condensate.}
% 	\label{fig:kdependence}
% \end{figure}
\begin{figure}[!thb]
%	\centering
	\includegraphics[width=0.45\textwidth]{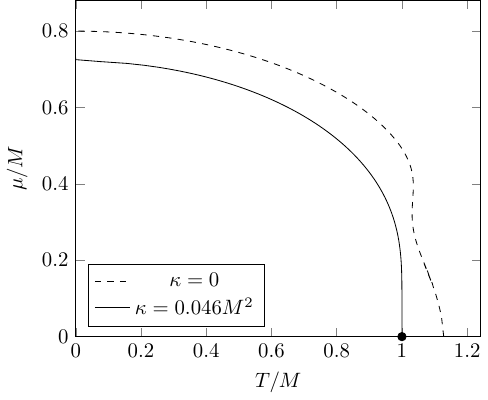}
	\caption{Phase diagram for the 2+1 dimensional Gross-Neveu model in the chiral limit  shows two phases, with nonzero and zero condensate, separated by a critical line for a second order transition in the chiral limit (solid line, $\kappa=0$) and a pseudocritical line for a crossover in the nonchiral case (dashed line, $\kappa=0.046~M^2$). 
    The black dot represents the point of instability.
    }
	\label{fig:critical_line}
\end{figure}

In the chiral limit $\kappa = 0$, the equation \eqref{eqn:gap_equation_2} 
{takes the form
\begin{equation}\label{eqn:critical_line}
	-M +\frac{1}{\beta}\ln[2+2\cosh(\beta \mu)] = 0,
\end{equation}
which describes a line that divides the temperature-chemical potential plane into two areas where $m$ is either zero or nonzero. }
Fig. \ref{fig:critical_line} shows the corresponding phase diagram in the $T-\mu$ plane. 
{The limiting values for critical temperature and chemical potential are $T_c(\mu=0)=M/2\ln 2$ and $\mu_c(T=0)=M$, respectively.} 

\begin{figure*}
\begin{subfigure}{0.4\textwidth}
    \centering
    \includegraphics[width=\linewidth]{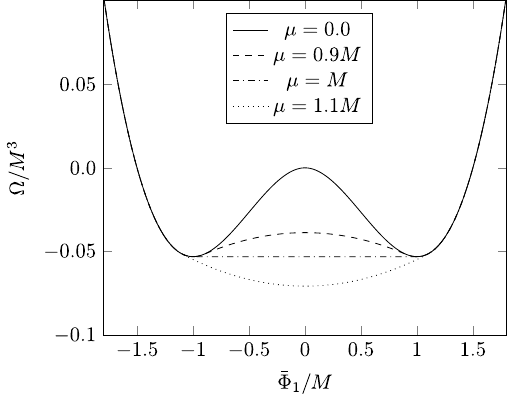}
    % \caption{Changes in the shape of phase shifts with external momentum.}
\end{subfigure}
\begin{subfigure}{0.4\textwidth}
    \centering
    \includegraphics[width=\linewidth]{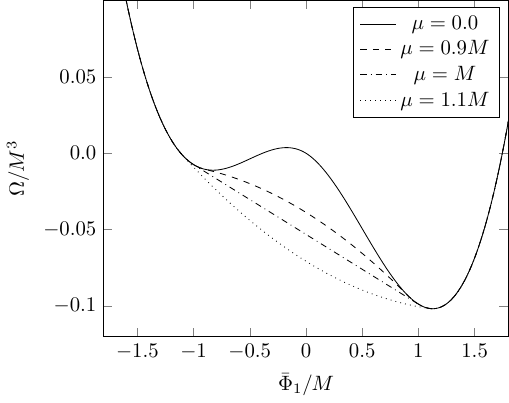}
    % \caption{Mott momenta as a function of temperature at $\mu=0.0$.}
\end{subfigure}
\caption{Grand potential $\Omega$ as a function of the order parameter $m$ 
{for different values of the chemical potential $\mu$ along a horizontal section on the phase diagram \ref{fig:critical_line} at $T=0.01~M$. 
The left panel shows the change in the shape of $\Omega$ for the chiral limit $\kappa=0$ and the right panel is for nonvanishing $\kappa=0.046$.}}
    \label{fig:order_phase_transition}
\end{figure*}

To gain insight into the order of the phase transition, we should look at the grand potential as a function of the order parameter $m$. Inside the area enclosed by the critical line and the axes in Fig.~\ref{fig:critical_line}, the thermodynamic potential is W-shaped and has two minima at finite values of $m$ 
(see Fig.~\ref{fig:order_phase_transition}), thus breaking the chiral symmetry. 
The W shape changes to a U shape with a minimum at zero while crossing the critical line. This hints at a second-order phase transition along this line. 
In the presence of non-zero $m_0$, the chiral symmetry is explicitly broken. 

In the zero-temperature limit, the grand potential behaves slightly differently. 
Using the identity %\eqref{eqn:identity1} 
\begin{equation}\label{eqn:identity1}
	\lim_{\beta\to\infty}\frac{1}{\beta}\ln\left (1+e^{\beta x}\right ) = x\theta(x)
\end{equation}
in Eq. \eqref{eqn:renorm_grand_potential}, the grand potential simplifies to
\begin{equation}
	\Omega(T\to0, \mu)/N = \begin{cases}
		\frac{m^2}{2g} + \frac{|m|^3}{3\pi}, & m >\mu \\
		-\frac{m^2}{2\pi}\left (\frac{\pi}{-g} - \mu\right ) - \frac{\mu^3}{6\pi}, & m < \mu .
	\end{cases}
\end{equation}
At $\mu = -{\pi}/{g} = M$ the grand potential admits minima for a wide range of $m$ values between $0$ and $M$. This instability gets enhanced when coupled to a magnetic field \cite{Lenz:2023gsq}.

In the case of graphene, this phase transition is known as semimetal-insulator transition \cite{Juricic:2009px}. This transition has been widely studied by various methods, e.g., the Dyson-Schwinger equation approach \cite{khveshchenko:2006} or lattice simulations \cite{Drut:2009aj}. Even though there have been many theoretical studies to understand this transition, it has not yet been observed experimentally. 
The paper \cite{liu:2009} attributed the inability to see this effect in experiments to the screening of the Coulomb interaction caused by disorder, doping, thermal effect, and volume effects, which strongly suppresses this transition even in the strong coupling regime.

{In the nonchiral case, the mass does not go to zero, and instead, we have a crossover. In figure \ref{fig:critical_line}, we show the pseudo-critical line along which the determinant 
\begin{equation}
    \begin{vmatrix}
        \frac{\partial^2 m}{\partial T^2} & \frac{\partial^2 m}{\partial T \partial \mu} \\
        \frac{\partial^2 m}{\partial \mu \partial T} & \frac{\partial m^2}{\partial \mu^2}
    \end{vmatrix}
\end{equation}
is zero. The line is shown for $\kappa=0.046M^2$ (see section \ref{sec:excitons} for the justification of such a choice). Other values of $\kappa$ show a similar behavior.
}

\begin{figure}
	\centering
	\includegraphics[width=0.5\textwidth]{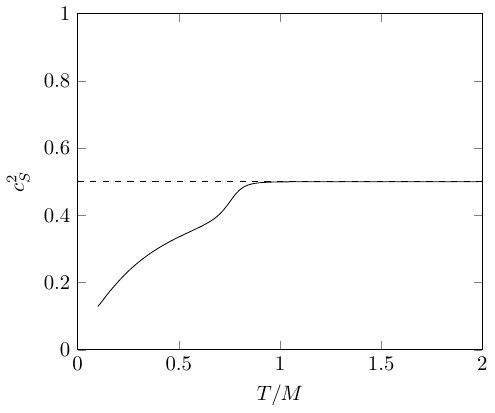}
	\caption{Speed of sound squared as a function of temperature. The dashed line shows the conformal limit $c_s^2 = 1/2$. The calculation is done at $\kappa=0.01$.}
	\label{fig:speed_of_sound}
\end{figure}

Thermodynamical quantities like pressure and energy can be calculated from the grand canonical potential.
For example, the pressure per particle species is $\mathcal{P} = -\Omega^{ren}/N - \mathcal{B}$, where the bag pressure term $\mathcal{B}$ is chosen so that the pressure of the vacuum ($T=\mu=0$) vanishes. The analytical expression for the pressure (in the $\Lambda \to \infty$ limit) is given below in terms of the polylogarithm function,
\begin{multline}
	\mathcal{P}(T,\mu) =  \dfrac{M}{2\pi}m^2 - \dfrac{m^3}{3\pi} - \dfrac{M^3}{6\pi} -  \dfrac{1}{\pi\beta^3}\left[\beta m \left(\textrm{Li}_{2}\left(-e^{-\beta(m - \mu)}\right) + \textrm{Li}_{2}\left(-e^{-\beta(m + \mu)}\right)\right) \right.\\ 
    \left. + \textrm{Li}_3\left(-e^{-\beta(m - \mu)}\right) + \textrm{Li}_3\left(-e^{-\beta(m + \mu)}\right)\right],
\end{multline}
where the mass $m$ is obtained by solving the gap equation \eqref{eqn:gap_equation_2}.
% For temperature larger than the critical temperature dynamical mass $\bar{\sigma}_1 = 0$,
% \begin{equation}
% 	\mathcal{P}(T,\mu) = -\dfrac{M^3}{6\pi} - \dfrac{1}{\pi\beta^3}\left[Li_3(-e^{-\beta \mu})+Li_3(-e^{\beta \mu})\right]
% \end{equation}
The entropy and the number density can be calculated from the relations 
$\mathcal{S} = {\partial\mathcal{P}}/{\partial T}$ and $\mathcal{N} = {\partial \mathcal{P}}/{\partial \mu}$, respectively. 
Using these relations, we can calculate the energy density
\begin{equation}
	\mathcal{E} = -\mathcal{P} + T\mathcal{S} + \mu\mathcal{N}.
\end{equation}
In Fig.~\ref{fig:speed_of_sound}, we have shown the squared speed of sound as a function of the temperature for $\mu=0$ in the medium calculated from the relation $c_s^2 = {d\mathcal{P}}/{d\mathcal{E}}$. 
The squared speed of sound correctly goes to the conformal limit of $1/2$ for high temperatures.
% add reference

% \begin{figure}[h!]
% 	\centering
% 	\begin{subfigure}{0.49\textwidth}
% 		\includegraphics[width=\textwidth]{plots/pressuredifferentk.png}
% 		\caption{Pressure}
% 		\label{fig:pres_k}
% 	\end{subfigure}
% 	\hfill
% 	\begin{subfigure}{0.49\textwidth}
% 		\includegraphics[width=\textwidth]{plots/energydifferentk.png}
% 		\caption{Energy}
% 		\label{fig:energy_k}
% 	\end{subfigure}
% 	\caption{Pressure and Energy at different $\kappa$}
% 	\label{fig:k_depen}
% \end{figure}

\section{Beyond Mean Field}
\label{sec:bu}

The mean-field approximation replaces the auxiliary fields with their expectation values. In doing so, it ignores the fluctuations effects of the auxiliary fields. In our method, the contribution of the fluctuation can be included by adding a perturbation to the mean fields $\bar{\Phi}_i \to \bar{\Phi}_i + \delta\Phi_i$. Substituting this into \eqref{eqn:partition_function}, inside the exponential we have the following terms,
\begin{equation}
	-N\int d^3x \sum_{i=1}^4\left ( \frac{\bar{\Phi}_i^2 + \delta\Phi_i^2 + 2\bar{\Phi}_i\delta\Phi_i}{2G} - \delta_{i1}\kappa(\bar{\Phi}_i + \delta\Phi_i) \right ) + N\rm{Tr}\ln S_{\rm tot}^{-1},
\end{equation}
where $S_{\rm tot}^{-1} = S_{\rm mf}^{-1} + \Sigma$ with $\Sigma = \sum_{i=1}^{4}\Gamma_i\delta\Phi_i$. The trace log part can be split into 
\begin{equation}
	\rm{Tr}\ln( S_{\rm tot}^{-1}) = \rm{Tr}\ln( S^{-1}_{\rm mf}) + \rm{Tr}\ln(1 + S_{mf}\Sigma).
\end{equation}
Now we can separate the mean field part as $\mathcal{Z} = \mathcal{Z}_{\rm mf}\mathcal{Z}_{\rm fl}$, with 
\begin{equation}
	\mathcal{Z}_{\rm fl} = \int\prod_{i=1}^4\mathcal{D}\delta\Phi_i \rm{exp}\left\{ -N\int d^3x \sum_{i=1}^4\left (\frac{\delta\Phi_i^2 + 2\bar{\Phi}_i\delta\Phi_i}{2G} - \delta_{k1}\kappa\delta\Phi_i\right ) + N\rm{Tr}\ln(1+S_{\rm mf}\Sigma)\right\}.
\end{equation}
The last term can be approximated as $\rm{Tr}\ln(1+S_{\rm mf}\Sigma) \approx  \rm Tr(S_{\rm mf}\Sigma) -\frac{1}{2} \rm{Tr}(S_{\rm mf}\Sigma S_{\rm mf}\Sigma)$. All the linear terms in field $\delta\Phi_i$ should vanish 
{as a consequence of fulfilling the gap equation so that} we are left with 
\begin{equation}
	\mathcal{Z}_{\rm fl} = \int\prod_{i=1}^4\mathcal{D}\delta\Phi_i \rm{exp}\left\{ -\int d^3x \sum_{i=1}^4\frac{\delta\Phi_i^2}{2}\left(\frac{N }{G} + N\rm{Tr}(S_{\rm mf}\Gamma_i S_{\rm mf}\Gamma_i)\right)\right\}.
\end{equation}
The second term inside the bracket looks similar to the polarization loop integral in the path integral. We define $\Pi_i = -N\rm{Tr}(S_{\rm mf}\Gamma_i S_{\rm mf}\Gamma_i)$. The expression of it is given below \cite{Ebert:2018dzs}

\begin{equation}\label{eqn:polarisation_function}
	\Pi_i(q,i\nu_m) = -N \sum_{\xi,\xi'=\pm1} \int^{|p|<\Lambda} \dfrac{d^2p}{(2\pi)^2}\dfrac{f^-(\xi'E_k) - f^-(\xi E_p)}{i\nu_m + \xi'E_k - \xi E_p}\left(1 -	\xi\xi'\dfrac{\vec{p}\cdot\vec{k} \mp m^2}{E_pE_k}\right),
\end{equation}
{where $\nu_m$ is the bosonic Matsubara frequency and the upper (lower) sign holds for $i=1,2 ~(3,4)$ }
%the fields $\Phi_{1,2}$ and the positive sign to the $\Phi_{3,4}$ 
and $\vec k = \vec p - \vec q$. 
Following the convention used in the NJL model, we will denote the fields $\Phi_{1,2}$ as scalar field $\sigma$ and $\Phi_{3,4}$ as pseudo-scalar field $\varphi$. 
By evaluating the Gaussian integral
{over the field fluctuations $\delta\Phi_i$, we find the relationship between the fluctuation contribution to the grand potential $\Omega_{\text{fl}}(T,\mu)$ and the polarization function,}
\begin{eqnarray}
\label{eqn:fluc_potential}
	\Omega_{\text{fl}}(T,\mu) &=& \frac{1}{2\beta l^2}\sum_{i=1}^4\ln \det \left[\mathcal{D}^{-1}_i(q,i\nu_m)\right]
    =\frac{1}{2\beta l^2}\sum_{i=1}^4 \rm{tr} \ln\left[\mathcal{D}_i^{-1}(q,i\nu_m)\right]
    , 
\end{eqnarray}
{where the latter enters the expression \eqref{eqn:fluc_potential} via the inverse propagator for the scalar and pseudoscalar fields, 
}
\begin{equation}
    \label{eq:propagator}
    \mathcal{D}^{-1}_i(q,i\nu_m) = \frac{N}{G} - \Pi_i(q,i\nu_m),
\end{equation}
and the "tr" symbol refers to summation in $(q,i\nu_m)$ space.
{In the following subsections, we will evaluate the polarization function and discuss the spectral properties of correlations beyond the mean field and their effect on the thermodynamics of the system.}
% Where for brevity, we write $\Pi_{\Phi_i}$ as $\Pi_i$ and so on.

%	or directly evaluating the formula
%	\begin{equation}
%		\textrm{Re}[\Pi_i(0,\omega + i\eta)] = -2N \dashint \dfrac{d^2p}{(2\pi)^2}[f^-(E_p) - f^-(-E_p)]\left[\dfrac{1}{\omega+2E_p} - \dfrac{1}{\omega - 2E_p} - \dfrac{1}{E_p}\right] \begin{pmatrix}p^2/E_p^2 \\ 1\end{pmatrix}
%	\end{equation}

\subsection{Bound Excitonic States} 
\label{sec:excitons}
First, we will focus on the case 
{of excitonic two-particle states at rest in the medium, i.e. we consider the polarization function for $q=0$. The zeros of the inverse propagator} 
$\mathcal{D}_i^{-1}$ at zero momentum correspond to the mass of bound exciton states,
\begin{equation}
	\frac{N}{G}  - \Pi_i(0,\omega = M_i+i\eta) = 0,
\end{equation}
where we have analytically continued the Matsubara frequencies to the real frequencies $i\nu_m \to \omega = M_i + i\eta$. 
When the mass of the excitonic state {exceeds twice the constituent mass, the bound state becomes unstable since the decay channel to its constituents opens (Mott dissociation).} 
In that case, we need to analytically continue the polarization function to the lower half of the complex $\omega$ plane \cite{Hufner:1996pq},

\begin{equation}
    \frac{N}{G}  - \Pi_i\left(0,\omega = M_i-i\Gamma_i/2\right) = 0,
\end{equation}
where the $\Gamma_i$ denotes here the decay width.

In this $q=0$ case, we have significant simplifications for the expression of the polarization function.
% \begin{equation}
% 	\Pi_i(0,i\nu_m) = -2N \int \dfrac{d^2p}{(2\pi)^2}[f^-(E_p) - f^-(-E_p)]\left[\dfrac{1}{i\nu_m+2E_p} - \dfrac{1}{i\nu_m - 2E_p}\right]\chi_i(p)	, \;  \chi_{1,2}(p) = \frac{p^2}{E_p^2} \text{ and } \chi_{3,4}(p) = 1\label{eqn:polarisation}
% \end{equation}
% Momentum and frequency independent part can be obtained by substituting $0$ in place of $\omega$ in the above equation.
% $$
% 	\Pi_{E,0} = -2N \int \dfrac{d^2p}{(2\pi)^2}\dfrac{f^-(E_p) - f^-(-E_p)}{E_p} \begin{pmatrix}p^2/E_p^2 \\ 1\end{pmatrix}
% $$
% It is interesting that the renormalization technique used at the mean-field level still works for this case.
%	$$
%		g^{-1} = G^{-1} - \dfrac{\Lambda}{\pi}
%	$$
Using the Sokhotski-Plemelj relation, 
\begin{equation}
\label{eq:plemelj}
   {\lim_{\eta\to 0}}~\frac{1}{x+i\eta} = \textrm{P.V.} \frac{1}{x} - i\pi\delta(x),
\end{equation}
and evaluating the {loop momentum integration in the polarization function} with the help of Dirac delta functions, we obtain the imaginary part,
\begin{equation}
	\textrm{Im} [\Pi_i(0,\omega+i\eta)] = \dfrac{N}{4} \omega(f^-(\omega/2) - f^-(-\omega/2))\chi_i(\omega)\Theta(\omega^2-4m^2)\Theta(4(\Lambda^2+m^2) - \omega^2), 
\end{equation}
where $\chi_{1,2}(\omega) = 1 - {4m^2}/{\omega^2}$, and $\chi_{3,4}(\omega) = 1$.
The real part can be calculated using the Kramers-Kronig relation, 
%or directly evaluating the principal value integral.
	\begin{equation}
    \label{eqn:kramers-kronig}
		\textrm{Re}[\Pi(q,\omega)] = \dfrac{1}{\pi}\dashint \dfrac{d\omega'}{\omega-\omega'}\textrm{Im}[\Pi(q,\omega')],
	\end{equation}
where the dashed integral symbol stands for the principal value integration. 
For the lower half-plane, the imaginary part does not have a simplified formula like the one above. 
So, we explicitly calculate the real and imaginary parts of $\mathcal{D}_i^{-1}=N/G  - \Pi_i$ and require both to be simultaneously zero. 
With these two equations, we can determine the two unknowns: the mass and the width for each channel. 

\begin{figure*}
\begin{subfigure}{0.45\textwidth}
	\centering
	\includegraphics[width=\textwidth]{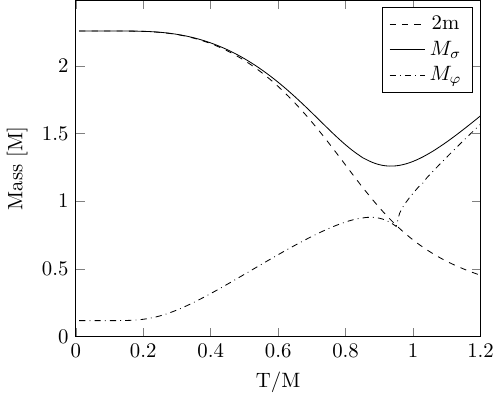}
\end{subfigure}
\begin{subfigure}{0.45\textwidth}
    \centering
	\includegraphics[width=\textwidth]{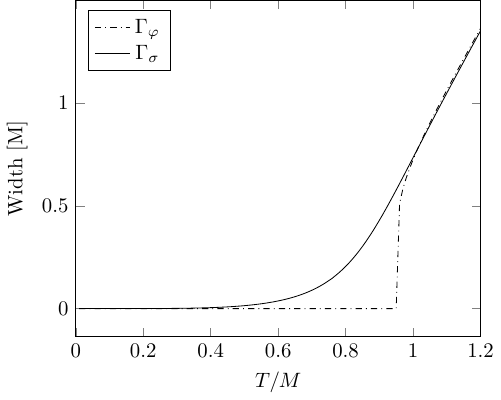}
\end{subfigure}
	\caption{Masses (left) and widths (right) of the excitonic states in the scalar (solid line) and pseudo-scalar (dash-dotted line) channels. The dashed line shows the twice of the condensate mass. The values are calculated at $\Lambda=5M$ and a small bare mass $m_0$ such that $M_\varphi(0,0)= 0.1M$.}
	\label{fig:masses}
\end{figure*}

The masses are shown in {the left and the decay width in the right panel of Fig. \ref{fig:masses}.
In local Gross-Neveu models, the scalar masses} are always heavier than twice the fermion mass, which makes them unstable\footnote{{In nonlocal extensions of the Gross-Neveu or Nambu--Jona-Lasinio (NJL) models, the masses of the scalar bound states are shifted below the two-particle decay threshold and thus are stable, as has been demonstrated in \cite{Schmidt:1994di}, see also \cite{Kachanovich:2016aun} for a recent application of this effect.}}. 
On the other hand, pseudo-scalar masses are much lighter and stable for small temperatures. Note that the pseudo-scalar particle corresponds to the Goldstone boson of the system. In the chiral limit ($m_0 = 0$), {the Goldstone boson should be massless according to the Goldstone theorem}. 
The pseudo-scalar states become unbound at the Mott temperature ($M_{\Phi_{3,4}}(T_{\rm Mott}) = 2m(T_{\rm Mott})$) and beyond. We also note that at high temperatures, both scalar and pseudo-scalar masses converge to each other. 
{This fact is a manifestation of the chiral symmetry restoration.}
In the left panel of Fig. \ref{fig:masses}, we show masses for a small value of $m_0$ at the cutoff of $5M$. 
{The Lagrangian mass parameter $m_0$ is chosen in such a way to have $M_\varphi = 0.1M$ at vacuum. 
Such a choice is made in analogy to the $3+1$ dimensional case of the NJL quark model where pion and/or kaon masses are fitted with experimental data.} 
With these parameters, we obtain the Mott temperature of $0.955\, M$. 
The rest of the figures below also use the same parameters.

In the case of graphene, this Mott criticality has been studied, e.g., in \cite{Herbut:2009vu, Classen:2015ssa}.

%		Note the Mott temperature decreases with the chemical potential.

\subsection{Phase Shift and Beth-Uhlenbeck approach}
The logarithm of the inverse propagator can be written in the spectral representation, 

\begin{equation}\label{eqn:phase_shift_defn}
    \frac{1}{2}\ln \left[\mathcal{D}_i^{-1}(q, i\nu_m)\right] = \int_{-\infty}^\infty \frac{d\omega'}{2\pi} \frac{\phi_i(q, \omega')}{i\nu_m - \omega'}.
\end{equation}
To find the spectral functions $\phi_i$ note that,
\begin{equation}
    \frac{1}{2}\ln\left[\frac{\mathcal{D}_i^{-1}(q, \omega + i\eta)}{\mathcal{D}_i^{-1}(q, \omega - i\eta)}\right] = \int_{-\infty}^\infty \frac{d\omega'}{2\pi} \phi_i(q, \omega') \left(\frac{1}{\omega - \omega' + i\eta} - \frac{1}{\omega - \omega' -i\eta}\right).
\end{equation}
Using the Sokhotski–Plemelj relation \eqref{eq:plemelj} on the last term, we find
\begin{equation}
    2\phi_i(q, \omega) = i \ln\left[\frac{\mathcal{D}_i^{-1}(q, \omega + i\eta)}{\mathcal{D}_i^{-1}(q, \omega - i\eta)}\right].
\end{equation}
In other words, $\phi_i(q, \omega) = \rm{Im}\ln \mathcal{D}_i^{-1}(q, \omega + i\eta)$. The above expression can be identified with the Jost representation of the S matrix $S_i(q, \omega) = {\mathcal{D}_i(q, \omega + i\eta)}/{\mathcal{D}_i(q, \omega- i\eta)}$, and assuming elastic scattering $|S_i| = 1$ we find a physical interpretation of the function $\phi_i(q, \omega)$ as the phase shifts, in particular $S_i(q, \omega) = e^{2i\phi_i(q, \omega)}$. 

\begin{figure}[!htb]
	\centering
	\includegraphics[width=0.9\textwidth]{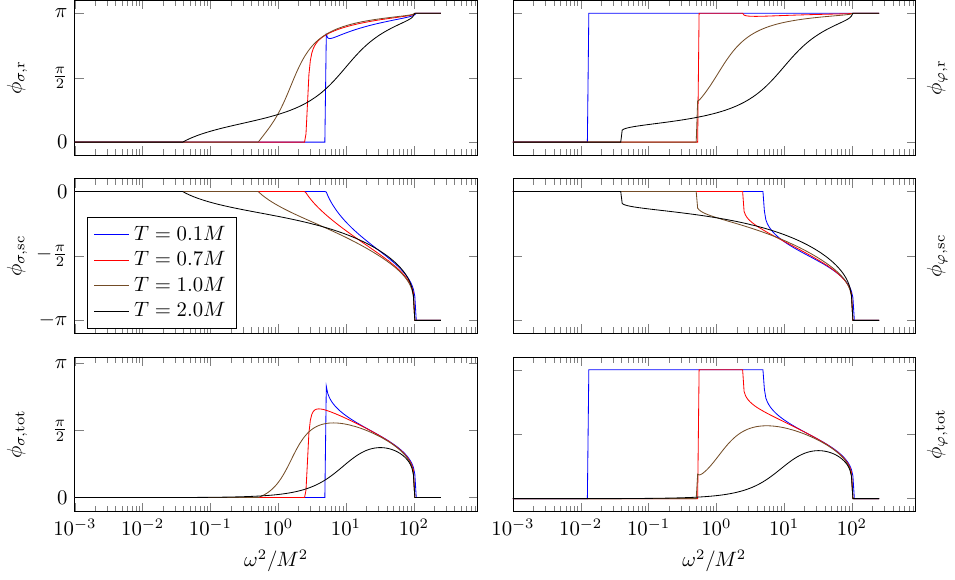}
	\caption{Phases for scalar (left) and pseudoscalar (right) channels. Resonant and scattering part of the phase are shown in the top and middle rows and the total phase is shown in the bottom row.}
	\label{fig:phase}
\end{figure}

As outlined in Ref.~\cite{Hufner:1994ma} and \cite{Wergieluk:2012gd}, the phase shift for fluctuations in the NJL model can be decomposed into a resonant part and a background contribution. 
Here, following Ref.~\cite{Blaschke:2013zaa}, we decompose the polarization function into a momentum and frequency-independent part $\Pi_{i,0}$ and the rest $\Pi_{i,2}(q,\omega+ i\eta)$,
% and the $\Pi_{0} = \Pi(0,0)$ is the part independent of external momentum and frequency and $\Pi_{2}$ is the residual part.
\begin{equation}
    	\mathcal{D}_i^{-1}(q,\omega+i\eta) = {N}/{G} - \Pi_{i, 0} - \Pi_{i, 2}(q,\omega+i\eta) = \Pi_{i, 2}(q,\omega + i\eta)(R_i(q,\omega+i\eta) - 1),
\end{equation}
with
\begin{equation}
	R_i(q,\omega+i\eta) = \dfrac{{N}/{G} - \Pi_{i, 0}}{\Pi_{i, 2}(q,\omega+i\eta)}.
\end{equation}
Introducing $R(q,\omega^2)$ in such a way enables us to write
\begin{eqnarray}
	\rm{Im} \ln \mathcal{D}_i^{-1}(q,\omega+i\eta) = \text{Im} \ln \Pi_{i, 2}(q,\omega + i\eta) + \text{Im} \ln[R_i(q,\omega + i\eta) -1] 
\end{eqnarray}
or equivalently
\begin{eqnarray}
\label{eq:phase}
\phi_i(q,\omega) = \phi_{i, \rm sc}(q,\omega) + \phi_{i, \rm R}(q,\omega),
\end{eqnarray}
where the phase shift is separated into the scattering part
\begin{equation}\label{eqn:phase_sc}
	\phi_{i, sc}(q,\omega) = - \arctan\left(\dfrac{\textrm{Im}\Pi_{i, 2}(q,\omega+i\eta)}{\textrm{Re}\Pi_{i, 2}(q,\omega+i\eta)}\right),
\end{equation}
and the resonant part
\begin{equation}\label{eqn:phase_R}
	\phi_{i, R}(q,\omega) = \arctan\left(\dfrac{\textrm{Im}(R_i(q,\omega^2))}{1 - \textrm{Re}(R_i(q,\omega^2))}\right).
\end{equation}

The phases at zero external momentum are shown in Fig.~\ref{fig:phase}. 
The figure is obtained by taking $\Lambda = 5M$.  The resonant part changing from $0$ to $\pi$ hints at the presence of bound states. The pseudoscalar channel shows this sharp jump below the Mott temperature. At higher temperatures, phase shifts for both channels become identical. All the phase shifts go to zero beyond the cutoff (at $\omega = 2\Lambda$). The qualitative form of the phase shift remains unchanged around the energy of the order $M$ as long as $\Lambda\gg M$.

\subsection{Effect of Finite Momenta}
In this section, we numerically explore the momentum-dependent part of the polarisation function.  
To calculate the imaginary part of the polarisation function with non-zero external momentum, we use the Sokhotski–Plemelj relation in \eqref{eqn:polarisation_function} to obtain the following expression.
	\begin{equation}
		\textrm{Im}[\Pi(q,\omega + i\eta)] = -N \sum_{\xi,\xi'=\pm 1}\int^{|p|<\Lambda} \dfrac{d^2p}{(2\pi)^2}(f^-(\xi' E_k) - f^-(\xi E_p)) (-\pi)\delta(\omega + \xi'E_k - \xi E_p)\mathcal{T}^{\mp}_{\xi \xi'}.
	\end{equation}
	where $\mathcal{T}^{\mp}_{\xi,\xi'} = \left(1 - \xi \xi'\frac{\vec p \cdot \vec k \mp m^2}{E_p E_k}\right)$ which can be simplified by assuming the constraint of the delta function to, $\mathcal{T}^{\mp}_{\xi \xi'} = \frac{-\xi \xi'}{2E_p E_k}(\omega^2 - q^2 - \psi_{\mp})$ with $\psi_{+} = 0$ and $\psi_{-} = 4m^2$. With this, the imaginary part has the following form.
	
	\begin{equation}
		\textrm{Im}[\Pi(q, \omega + i\eta)] = N(s - \psi_{\mp})\left [ -\theta(-s)\int_{y}^{\sqrt{\Lambda^2 + \bar{\Phi}_1^2}}dx\frac{B_1(x)}{F(x)} + \theta(s-4m^2)\int_{-y}^{y}dx\frac{B_2(x)}{F(x)}\right ].
	\end{equation}
	where the Pauli blocking factors are,
	\begin{equation}
		B_2(x) = f^-\left(\frac{-x - \omega}{2}\right) - f^-\left(\frac{-x + \omega}{2}\right), \qquad B_1(x) = B_2(x) + B_2(-x),
	\end{equation}
	and the factor in the denominator is
	\begin{equation}
		F(x) = 4\pi\sqrt{q^2(s - 4m^2) - s x^2},
	\end{equation}
	and $y = q\sqrt{1 - 4m^2/s}$.
	
	The real part can be calculated using the Kramers-Kronig relation \eqref{eqn:kramers-kronig}.

Before presenting the momentum dependence of the phase shifts, it is worthwhile to note that even though the excitonic states will gain a momentum dependence with respect to the rest frame of the medium, the constituent mass, as calculated earlier, remains unaffected by it. For self-consistency, we should also include momentum dependence in the mass gap. The paper { \cite{Blaschke:1994px} discusses this in more detail, with momentum distributions for pairs moving in the medium that have also been used, e.g., in \cite{Barz:1992ra}.}
We include the momentum dependence on the mass gap by rewriting the equation \eqref{eqn:gap_equation} as

    \begin{equation}
        \frac{\partial \Omega_{\rm{mf, vac}}^{\rm{ren}}}{\partial m(q)} + \int^{|p|<\Lambda} \frac{d^2 p}{(2\pi)^2} \frac{2m(q)}{E_p}\left(f\left(E_{\mathbf{p}+\mathbf{q}/2} + \mu\right) + f\left(E_{\mathbf{p}-\mathbf{q}/2} - \mu\right)\right) = 0.
    \end{equation}

\begin{figure}[thb]
        \centering
        \includegraphics[width=0.5\linewidth]{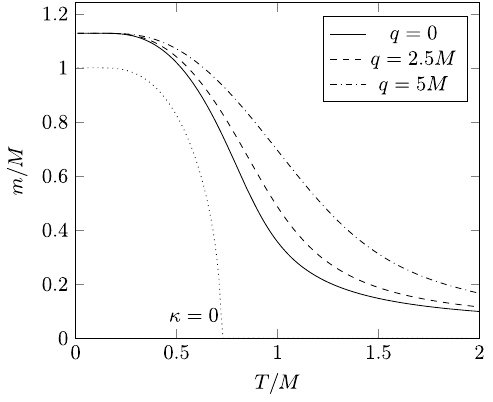}
        \caption{{Temperature dependence of the mass gap for fermions in a pair moving with center of mass momentum $q$ relative to the medium.} The dotted line shows the mass gap in the chiral limit {for zero momentum, whereas the solid, dashed, and dash-dotted lines are for different momenta $q$ } at a finite value of $\kappa=0.046$.}
        \label{fig:mass_gap_q}
    \end{figure}

Figure \ref{fig:mass_gap_q} shows 
{the temperature dependence of the fermion mass gap for different center of mass momenta of the pair boosted relative to the rest frame of the medium.} 
All the calculations below use this momentum-dependent mass gap.

\begin{figure}[htb]
	\centering
	\includegraphics[width=0.6\textwidth]{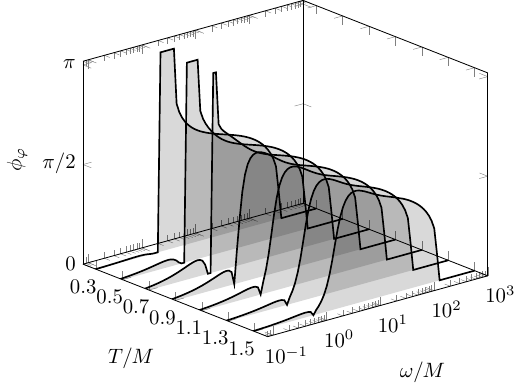}
	\caption{Phase shifts for the pseudo-scalar channel at external momentum $q = M$, shown at various temperatures. The phase shift gains a bulge in the region $s<4m^2$ for temperature near Mott temperature.}
	\label{fig:phase_momentum_dependence}
\end{figure}

The {scattering and resonant contributions to the} phase shifts \eqref{eq:phase} are then obtained using \eqref{eqn:phase_sc} and \eqref{eqn:phase_R}, respectively, and are shown in the figure \ref{fig:phase_momentum_dependence} 
{for the finite momentum $q=M$. 
These phase shifts are very similar to those at rest,} 
but there are two major differences. 
First, at finite momenta, 
{there are non-zero values of the phase shifts in}
the region $s=\omega^2 - q^2 <0$, 
{which correspond to so-called Landau damping modes} \cite{Maslov:2023boq,Ghosh.etal:2024}. 
This holds for both scalar and pseudo-scalar channels. 
Figure \ref{fig:phase_momentum_dependence} shows this effect for the pseudo-scalar channel at the specific momentum ($q=M$). The small inflation in the region $s<0$ can be clearly seen. 

\begin{figure*}
\begin{subfigure}{0.45\textwidth}
    \centering
    \includegraphics[width=\linewidth]{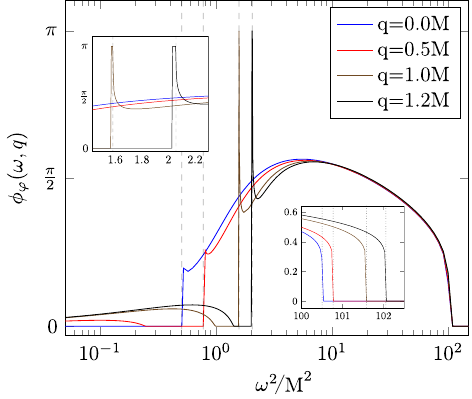}
    % \caption{Changes in the shape of phase shifts with external momentum.}
    \label{fig:boost}
\end{subfigure}
\begin{subfigure}{0.45\textwidth}
    \centering
    \includegraphics[width=\linewidth]{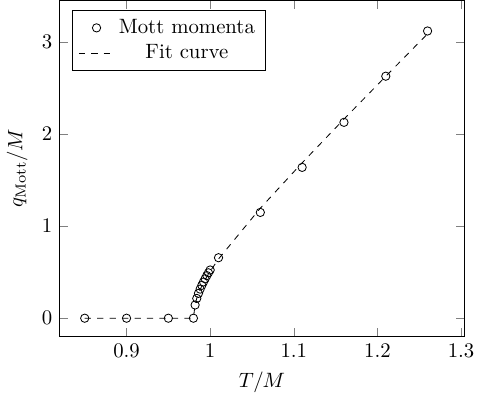}
    % \caption{Mott momenta as a function of temperature at $\mu=0.0$.}
    \label{fig:mott_momenta}
\end{subfigure}
\caption{Effect of external momenta on the phase shifts. The left panel shows the change in the shape of the phase shift for the pseudo-scalar channel at a temperature above the Mott temperature (the figure is at $T = M$). The figure shows that bound states reappear at high values of external momenta. The dashed lines correspond to the threshold values $\omega^2 = q^2 + 4m^2$ and the dotted lines (in the inset) are the maximum frequencies allowed by the cutoff $\omega_{\rm{max}}^2 = 4\Lambda^2 + q^2 + 4m^2 $. The right panel shows the Mott momentum $q_{\rm Mott}$ as a function of temperature (open dots) which can be well fitted with the simple formula \eqref{eqn:fit-formula} (dashed line).}
\label{fig:Mott_momenta}
\end{figure*}

{Second,} the pseudo-scalar channel exhibits another intriguing phenomenon,
{the resurrection of the bound state for sufficiently large momenta above the Mott temperature $T_{\textrm{Mott}}=0.98~M$}. 
For bound states at rest ($q=0$), below the Mott temperature, the phase shift shows a sharp jump from $0$ to $\pi$ {at the position of the bound state, stays at $\pi$ until the continuum threshold at $\omega=2m$ is reached. From there, the phase shift decreases and reaches zero at $\omega_{\rm{max}}^2 = 4\Lambda^2 + q^2 + 4m^2 $, where it stays for all $\omega$ up to infinity. 
However, at finite temperatures the continuum threshold is lowered towards chiral symmetry restoration and consequently has to cross the mass of the exciton which therefore becomes unbound. This situation defines the Mott temperature.
According to Levinson's theorem, the vanishing of the bound state is reflected in the behavior of the phase shift at the threshold by a jump from $\pi$ to zero.
Remarkably, for temperatures above the Mott temperature, one can observe the restoration of this bound state for exciton momenta exceeding a certain value: the Mott momentum $q_{\rm Mott}$. 
This effect occurs because the Pauli blocking of the electron-hole interaction gets reduced as a function of the momentum which boosts the electron-hole pair relative to the surrounding medium.
A similar effect has been discussed for deuterons in nuclear matter in Ref. \cite{Ropke:1983lbc}.

The left panel of Figure \ref{fig:Mott_momenta} shows this transition in the shape of the phase shift from a smooth function that stays well below $\pi$ to one which shows a sharp jump by $\pi$ when the Mott momentum is exceeded, signalling} the resurrection of the bound state. 
{Both effects of a finite momentum, the appearance of Landau damping modes and the resurrection of a bound state beyond the Mott momentum, demonstrate that the phase shifts in medium are not invariant against a Lorentz-boost. This was to be expected because the rest frame of the medium represents a special distinguished frame of reference, see also \cite{Benic:2013eqa} for Lorentz symmetry breaking effects in a Nambu--Jona-Lasinio model. }

On the right panel of Figure \ref{fig:Mott_momenta}, we show the smallest external momentum needed for this jump to happen, denoted as the Mott momentum $q_{\textrm{Mott}}$, as a function of the temperature. The Mott momenta are fitted with the formula

\begin{equation}\label{eqn:fit-formula}
    q_{\textrm{Mott}} = 9\sqrt{(T-T^*)^2 - (T_{\textrm{Mott}}-T^*)^2}, \qquad \text{for }
    T\ge T_{\textrm{Mott}}, \text{ with } T^* = 0.91~M.
\end{equation}
%The Mott momenta follows a very simple formula of $q_{\rm Mott} = \sqrt{2(T^2 - T_{\rm Mott}^2)}$ where $T_{\rm Mott}$ is the Mott temperature.
% In fact the total pressure will be slightly lowered leading to the effect of Landau damping.
{As an intuitive explanation, one can think of an analogon to the photoelectric effect. The exciton that is immersed in the surrounding medium can become apparent as a sharp bound state peak in the spectrum with the dispersion relation $E(q)=\sqrt{q^2+{m^*}^2}$ and an effective mass $m^*=9(T_{\rm Mott}-T^*)=0.63\, M$ for $q\ge q_{\rm Mott}$, when the thermal energy $E_{\rm th}=9T$ of an impacting quasiparticle exceeds this exciton energy by the amount of a "binding energy" $E_b=9T^*=8.19\, M$. }

\subsection{Pressure of excitonic fluctuations}

In addition to giving an intuitive physical picture 
{and a relation to observable processes, the phase shifts allow to quantify the contribution of the continuum of scattering states to the thermodynamic quantities in the spirit of the work by Beth and Uhlenbeck \cite{Beth:1937zz}. }
For example, the pressure due to fluctuations can be calculated from the generalized Beth-Uhlenbeck formula \cite{Blaschke:2013zaa},
\begin{equation}\label{eqn:pressure_fluc}
    \mathcal{P}_{\rm{fl}} = \sum_{i}\int^{|p|<\Lambda} \frac{d^2q}{(2\pi)^2}\int_{-\infty}^{\infty}\frac{d\omega}{2\pi}g(\omega)\phi_i(q, \omega).
\end{equation}
where $g(\omega) = (e^{\beta \omega} - 1)^{-1}$ is the Bose-Einstein distribution function. The formula can be obtained by substituting \eqref{eqn:phase_shift_defn} into \eqref{eqn:fluc_potential} and evaluating the Matsubara sum.

% \cite{Ebert:2018dzs} the author considered a pole approximation,

% \begin{equation}
%     \frac{d\phi_i}{d\omega} \approx \pi\delta(\omega  - E_i)
% \end{equation}
% with $E_i = \sqrt{q^2 + M_i^2}$. Whereas in 
In Refs. \cite{Ebert:2018dzs} and \cite{Blaschke:2013zaa}, the authors have considered the approximation that phase shifts are Lorentz boost invariant. In this scheme, one performs the calculation of the phase shift at rest and obtains the momentum-dependent phase shifts by a Lorentz boost from the rest system, $\phi_i(q, \omega) = \phi_i(q=0, \sqrt{\omega^2 - q^2})$, with $\phi_i = 0$ for the case of $\omega < q$.

\begin{figure}
    \centering
    \includegraphics[width=0.7\linewidth]{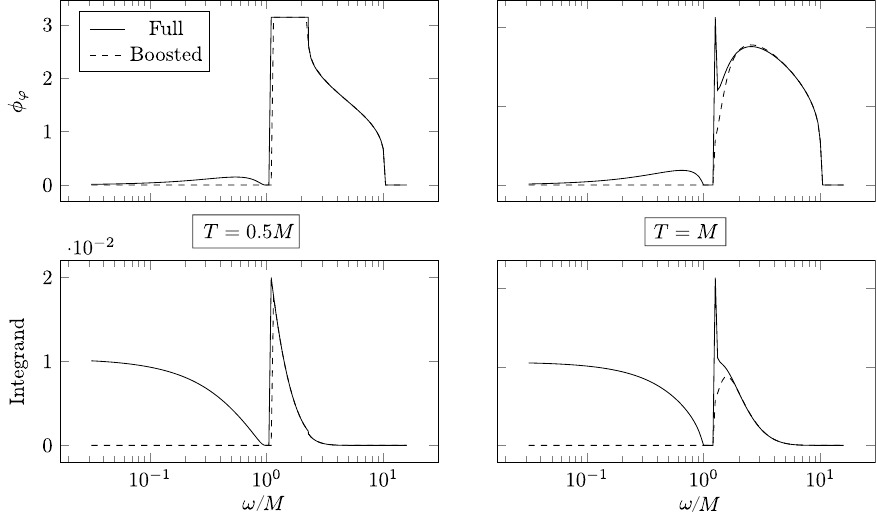}
    \caption{Comparison of the phase shifts 
    {(upper panels) and the contributions to the pressure (lower panels) for the full calculation (solid lines) and the boost approximation (dashed lines) at a temperature above (right panels) and below (left panels) the Mott temperature for the momentum $q=M$.}}
    \label{fig:integrand_comparison}
\end{figure}

Figure \ref{fig:integrand_comparison} compares the phase shifts calculated at finite momentum $q=M$ with those obtained by Lorentz boost from the $q=0$ ones for two different values of the temperature. The main difference is in the Landau region and in the region near the threshold where the bound state reappears. The bottom panel shows the integrand of the pressure \eqref{eqn:pressure_fluc}, i.e. the phase shift multiplied with the Bose function. The Landau region gives the dominant contribution to the pressure. A similar result without the resurrection of the bound state could also be seen in the case of the scalar channel.

\begin{figure*}
\begin{subfigure}{0.45\textwidth}
    \centering
    \includegraphics[width=\linewidth]{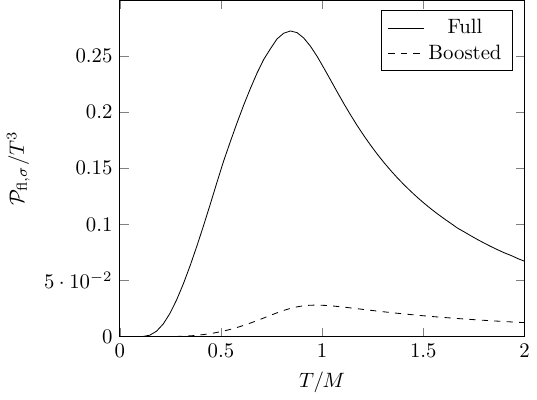}
    % \caption{Changes in the shape of phase shifts with external momentum.}
    \label{fig:pressure_comparison_sigma}
\end{subfigure}
\begin{subfigure}{0.45\textwidth}
    \centering
    \includegraphics[width=\linewidth]{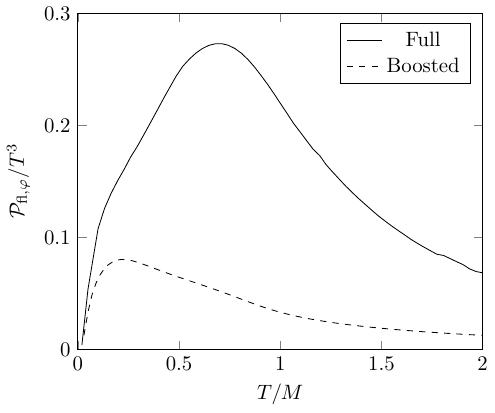}
    % \caption{Mott momenta as a function of temperature at $\mu=0.0$.}
    \label{fig:pressure_comparison_phi}
\end{subfigure}
\caption{Comparison of fluctuation pressure as a function of the temperature for scalar (left) and pseudo-scalar (right) channels. The full line corresponds to the full momentum-dependent calculation of the pressure, whereas the dashed and dotted lines correspond to the boost approximation and the pole approximation, respectively.}
\label{fig:pressure_comparison}
\end{figure*}

In figure \ref{fig:pressure_comparison}, we compare the pressure in the boost approximation with the full momentum-dependent calculation. Due to the Bose-Einstein factor multiplying the phase shift, the minute difference in the $s<0$ case gets amplified, and we have a significant increase in the fluctuation pressure. 

\begin{figure}
    \centering
    \includegraphics[width=0.6\linewidth]{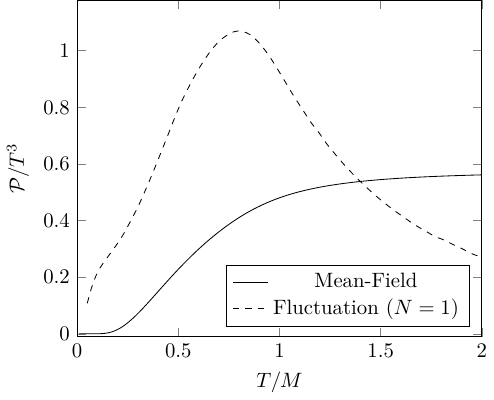}
    \caption{Comparison of mean field pressure with the fluctuation pressure as a function of temperature. The fluctuation pressure is shown at $N=1$ species of Fermionic fields.}
    \label{fig:press_comp_mean_field_fluc}
\end{figure}

{Finally, in Figure \ref{fig:press_comp_mean_field_fluc}, we compare the total fluctuation pressure with the mean-field level.} 

Similar calculations for the case of 3+1 dimensions can be found in \cite{Maslov:2023boq} for the NJL model coupled to the Polyakov loop.

\section{Conclusions}
In this article, we have explored a graphene-inspired generalized Gross-Neveu model in $2+1$ dimensions. At the mean-field level, the model shows a second-order transition from a semimetal to an insulator with an instability at the point $T=0, \mu = M$. This instability gets enhanced when the theory is coupled to an external magnetic field \cite{Lenz:2023gsq}. When compared to the 3D counterpart of the model, the thermodynamical quantities at the mean field have a different scale dependence on temperature and chemical potential but show a similar trend when normalized appropriately. Beyond the mean field, the bound state spectra and the phase shifts follow a similar pattern to the corresponding 3D version. The difference lies in the extent of renormalizability. It shows up in many ways, from having less dependence on the cutoff to having the speed of sound subluminal. The main crux of the papers deals with the momentum dependence of the phase shifts and its implication on thermodynamics. Here, we presented a formula in terms of integrals for the imaginary part of the polarization loop. With this full momentum-dependent phase shift calculation, we see a huge discrepancy in pressure from the previously used boost assumption. This difference can be attributed to two critical phenomena: the existence of a non-zero phase shift in the Landau region ($s<0$) and the resurrection of the bound state for the pseudo-scalar channel above the Mott momentum. 

%Here we allude to the difference between the existing method of using boosts from the phase shifts at rest and the full momentum-dependent calculations. The actual difference and how it affects the fluctuation pressure etc. will be of particular importance and are delegated to the forthcoming publication.

\begin{acknowledgments}
B.M. acknowledges a stipend from the International Max Planck Research School for Quantum Dynamics and Control at the Max-Planck Institute for Physics of Complex Systems in Dresden, Germany. The work of D.B. and B.M. was supported by NCN under grant No. 2021/43/P/ST2/03319.
\end{acknowledgments}

\appendix
\section{Gamma Matrices} \label{appendix:gamma}
The following irreducible representation of gamma matrices is used in this text,
	\begin{equation*}
		\gamma_k = \begin{pmatrix}
			0 & i\tau_k \\
			-i\tau_k & 0
		\end{pmatrix}
		\text{,} \qquad
		\gamma_3 = \begin{pmatrix}
			0 & I_2 \\
			I_2 & 0
		\end{pmatrix},
	\end{equation*}
 There exist two other $4\times 4$ matrices that anti-commute with the above three and with each other,
	\begin{equation}
		\gamma_4 = \begin{pmatrix}
			0 & i\tau_3 \\
			-i\tau_3 & 0 
		\end{pmatrix}
		\text{,} \qquad
		\gamma_5 = \gamma_1\gamma_2\gamma_3\gamma_4 = \begin{pmatrix}
			I_2 & 0 \\
			0 & I_2
		\end{pmatrix}.
	\end{equation}
The matrix $\gamma_{45}$ which commutes with $\gamma_\mu$ but anti-comutes with $\gamma_4$ and $\gamma_5$ is then defined by 
 \begin{equation}
     \gamma_{45} = \frac{i}{2}[\gamma_4, \gamma_5] = \begin{pmatrix}
         0 & \tau_3 \\
         \tau_3 & 0
     \end{pmatrix}.
 \end{equation}

\section{Regulator Dependence} \label{appendix:regulator}
The divergent part in the grand potential \eqref{eqn:omega} is $\int \frac{d^2p}{(2\pi)^2} E_p^{(k)}$. Introducing a sharp momentum cutoff as a regulator, we have
\begin{equation}
     \int_0^\Lambda \frac{pdp}{2\pi} E_p^{(k)} = \frac{(\Lambda^2 + M_k^2)^{3/2} - |M_k|^3}{6\pi}.
\end{equation}
In our case, $M_k = \bar{\Phi}_1 = m$, so the summing over two different $k$'s results in $[(\Lambda^2 + m^2)^{3/2} - m^3]/(3\pi)$. Expanding this in powers of $m/\Lambda$, we get the first term of a series,
\begin{equation}
     - \frac{|m|^3}{3\pi} + \frac{\Lambda^3}{3\pi} + \frac{\Lambda m^2}{2\pi} - \frac{m^4}{8\pi\Lambda} + \cdots .
\end{equation}
The constant term $\Lambda^3/3\pi$ can be removed by shifting the potential. The term proportional to the regulator we have absorbed into the coupling by introducing the renormalized coupling $g$ in the main text, see Eq. \eqref{eqn:g}. In the large cutoff limit, the last term should go to zero. However, in the case of a finite cutoff, this term plays quite an important role. 
The grand potential can be expanded analytically to have the form $\Omega = a_0 + a_2 m^2 + a_4 m^4 + a_6 m^6 + \cdots$, where 
\begin{align}
    a_0 &= \left(\textrm{Li}_3(-e^{\beta \mu}) + \textrm{Li}_3(-e^{-\beta\mu})\right)/\pi\beta^3, \\
    a_2 &= \left(-M + T [\log(1 + e^{-\beta\mu}) + \log(1 + e^{\beta\mu})] \right)/2\pi,\\
    a_4 & = \frac{\beta e^{\beta \mu}}{4\pi(1 + e^{\beta\mu})^2} - \frac{1}{8\pi\Lambda}.
\end{align}
In the continuum case ($\Lambda\to\infty$), the term $a_4$ is always positive. Then, for positive $a_2$, we have only one minimum in the potential at $m=0$, while for negative $a_2$, there is a maximum at origin and two minima at finite values of $m$ and $-m$. The phase transition corresponding to this is second order, and the line of $a_2=0$ (same equation as in \eqref{eqn:critical_line}) separates the two phases.

\begin{figure*}
\begin{subfigure}{0.45\textwidth}
    \centering
    % \caption{Changes in the shape of phase shifts with external momentum.}
    \includegraphics[width=\linewidth]{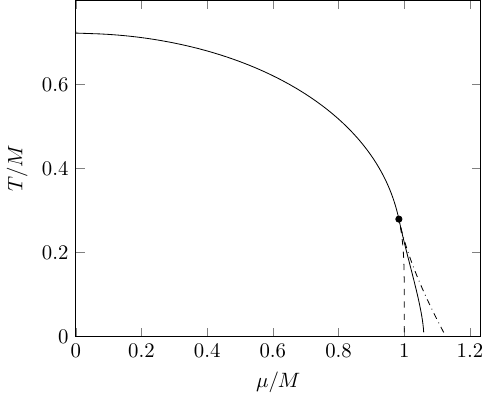}
    \label{fig:phase_diagram_regulator_dependence}
\end{subfigure}
\begin{subfigure}{0.45\textwidth}
    \centering
    \includegraphics[width=\linewidth]{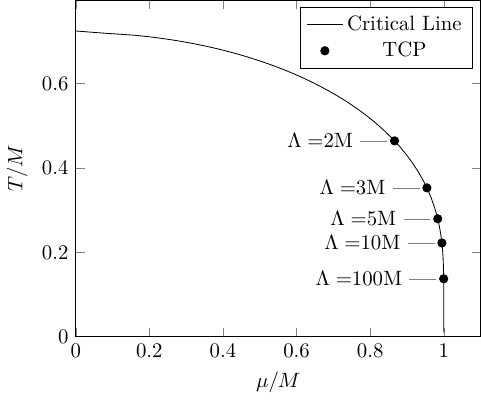}
    % \caption{Mott momenta as a function of temperature at $\mu=0.0$.}
    \label{fig:phase_diagram_cutoff}
\end{subfigure}
\caption{Phase Diagram at a finite value of the regulator. On the left, we show the phase diagram at $\Lambda=5M$, and on the right, we show the tricritical points for different values of the regulator. In the phase diagram on the left below the tri-critical point (TCP), we have a first-order transition. The solid line separates regions with zero and non-zero values of $m$. The region inside the dashed and dash-dotted line corresponds to the mixed phase. On the right, we show the position of the TCP on the $a_2=0$ line for different values of the cutoff.}
\label{fig:phase_diagram_regulator}
\end{figure*}

In the finite regulator case, the presence of the last term in the expression of $a_4$ makes it possible to have negative values. 
When $a_2>0$ and $a_4$ becomes negative, we have one minimum at $m=0$ and two other minima at finite $m$. The global minimum changes from one to the other, giving rise to a first-order phase transition.  On the right panel of Fig.~\ref{fig:phase_diagram_regulator} we show the line for $a_2=0$ as well as the points below which $a_4$ turns negative for different values of cutoffs. 
On the left panel, we show the phase structure for a particular cutoff value of  $\Lambda=5~M$. Consider two separate temperature slices above and below the tricritical point(TCP) (shown by a black dot). 
For temperatures above the TCP, the grand potential changes shape from W to U as one increases chemical potential, typical of a second-order phase transition. For the case of temperature lower than TCP, The W shape changes to a shape with three minima after crossing the dashed line and then changes to a U shape after crossing the dash-dotted line. In the region bounded by the TCP, dashed, dash-dotted lines, and the $T=0$ line, we have three minima. 
Along the solid line below the TCP, all minima are degenerate. This situation corresponds to the coexistence of the massless and massive phases and thus indicates the existence of a mixed phase. In the region between this solid line and the dashed line, the global minima are at finite $m$ and the region between the solid and the dash-dotted line the global minimum is at $m=0$. Ref.~\cite{Buballa:2020nsi} presented similar results using the Pauli-Villars regularization scheme.
 
\bibliography{ref}
\end{document}